\documentstyle[pra,eqsecnum,aps,epsfig]{revtex}

\begin{document}

\title{Theory of strong outcoupling from Bose-Einstein condensates}
\author{Robert Graham\thanks{Permanent address:
Fachbereich Physik, Universit\"at GH Essen, 45117 Essen,
Germany} and Dan F.~Walls}
\address{
Department of Physics\\University of Auckland\\
Private Bag 92019\\ Auckland, New Zealand}
\maketitle

\begin{abstract}
We study the dynamics of a magnetically trapped Bose-Einstein condensate
in the presence of an external electromagnetic field coupling trapped,
untrapped and antitrapped Zeeman sublevels. For large condensates an
approximate analytical solution of the coupled Gross-Pitaevskii equations
is given in the regime of strong outcoupling. The theory is developed for
the cases of rf-outcoupling within a hyperfine manifold of states, micowave-outcoupling connecting states in two different hyperfine manifolds, and Raman outcoupling.
\end{abstract}

\section{Introduction}\label{sec:1}

One of the most promising future applications of the Bose-Einstein condensates
of magnetically trapped atoms is the possible realization of efficient atom
lasers, whose experimental proof of principle has already been given \cite{1}.
An essential element of any atom laser is the process of outcoupling from
the trapped Bose-Einstein condensate. In \cite{1} an $rf$-field was used to
coherently transfer atoms from the trapped state to untrapped magnetic
sublevels within the same hyperfine manifold of states, enabling them to leave
the trap under the influence of gravity. Another outcoupling scheme presently
being explored experimentally \cite{2} is based on Raman transitions in an
external optical field.

The theory of outcoupling from Bose-Einstein condensates has been developed
in a number of papers \cite{3,3a,HMS,5,5a}. The theory is based on the
Gross-Pitaevskii equations for the macroscopic wave functions of the atoms in
the trapped and untrapped magnetic sublevels, coherently coupled by the
externally imposed electromagnetic field. In \cite{3} the coupled
Gross-Pitaevskii equations for two magnetic sublevels were solved numerically
in 1 dimension for a number of cases, and various dynamical regimes
were identified. In \cite{3a} a detailed comparison of numerical and
experimental results was given. In \cite{4} the regime of weak outcoupling was
considered and in this limit an approximate analytical solution of essentially
the same theoretical model as in \cite{3} and \cite{3a} was given. The
weak outcoupling regime is defined physically by the condition that the Rabi
oscillations induced by the electromagnetic fields in a certain
resonance zone within the condensate are slow on the time scale on which the
driven atoms leave that zone. An experimental study \cite{6} of this regime was recently performed. In this limit the outcoupling proceeds
at an approximately time-independent perturbatively calculable rate \cite{4},
and the reduced effective condensate dynamics can be described by a Markoff
process, similar to the familiar Markovian description of a lossy laser mode.
The opposite regime of strong outcoupling, where the atoms move very slowly
on the time-scale of the Rabi cycles, has also been seen in numerical
simulations \cite{3,4} but has not yet been studied theoretically in detail.
In fact, the experimental realization of an atom laser in \cite{1} operated
in this regime, which is therefore of practical as well as theoretical
interest. In the present paper we give an approximate analytical solution of
the coupled   Gross-Pitaevskii equation in the regime of strong outcoupling.
This will be done by a method similar to that of \cite{7},\cite{8} for the solution of the time-dependent Gross-Pitaevskii equation,

\section{Model of the outcoupling process}\label{sec:2}

The model of the outcoupling process we shall analyse in this paper is
essentially that of ref. \cite{4}, but with allowance for the action of
gravity, which is actually very important \cite{5}, at least in the case of microwave
or $rf$-field outcoupling, for extracting the outcoupled atoms from the trap. Thus we consider a coupled set of Gross-Pitaevskii
equations for the macroscopic wave function $\psi_m$ in the different
internal atomic states, e.g. Zeeman sublevels, labelled by $m$. We shall
assume that the interaction energy between the atoms is a functional of the
total density of the atoms $|\psi|^2=\sum_m|\psi_m|^2$ only and not dependent
on the internal state $m$. This assumption is rather well satisfied within
a hyperfine manifold of fixed total spin quantum number $F$, but we shall use
it as a simple model assumption also for levels with different $F$-values.
Then the coupled Gross-Pitaevskii equations for $rf$-outcoupling from an $F=1$ hyperfine manifold or microwave oucoupling between two different hyperfine states take the form:
\begin{equation}
i\hbar\dot{\psi}_m
=
\left(-\frac{\hbar^2\nabla^2}{2M}+V_m(\bbox{x})+U_0|\psi|^2\right)\psi_m
+\hbar\Omega^* e^{i\omega t}\psi_{m+1}+ \hbar\Omega e^{-i\omega t}\psi_{m-1}
\label{eq:2.1}
\end{equation}
$M$ is the atomic mass, $U_0=\frac{4\pi\hbar^2a}{M}$ describes the effective
low-energy scattering potential with scattering length $a$. We shall restrict
our discussion to the case of repulsive interaction $a>0$.

\begin{figure}
\centerline{
\epsfig{figure=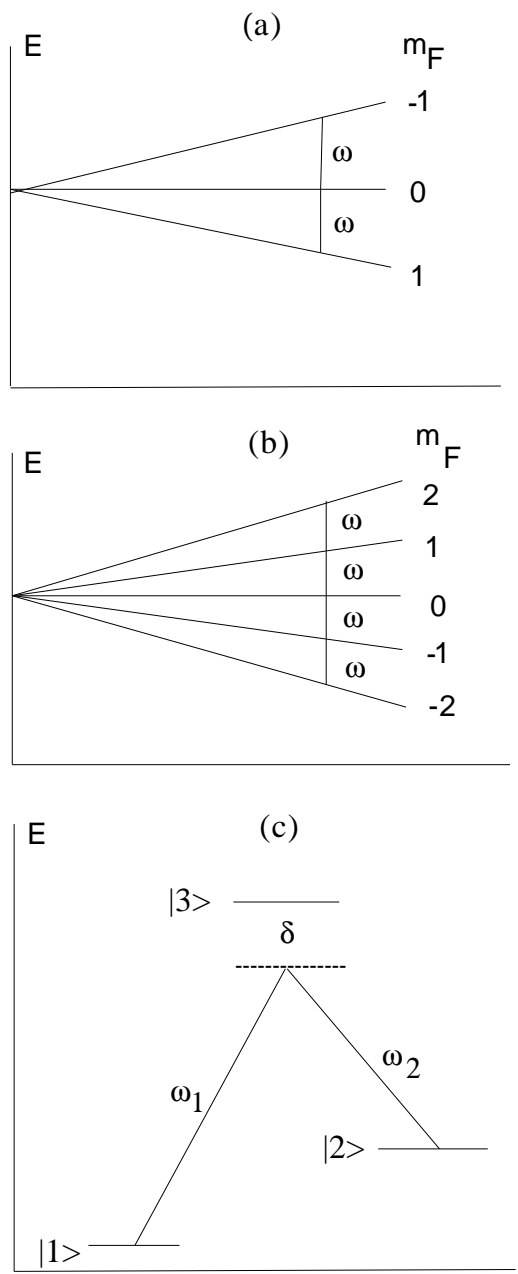}}
\caption{
Outcoupling schemes for radio-frequency fields (a) micro-wave fields (b) and
optical fields (c)}\label{1}
\end{figure}

For $rf$-outcoupling from an $F=1$, $m_F=-1$ condensate, like the $^{23}Na$
condensate in \cite{1} (cf. fig.\ref{1}(a)), $m=-1,0,1$ labels the 3 magnetic
sublevels of the $F=1$ state with trapping potentials
\begin{eqnarray}
V_{\pm1} &=&
\pm\left(V(0)+\frac{M}{2}\bar{\omega}^2r^2\right)+Mg(z-z_0)\\
\label{eq:2.2}
V_0 &=& Mg(z-z_0)\,.\nonumber
\end{eqnarray}
Here and in the following we neglect quadratic Zeeman shifts. Thus $m=0$ and $m=1$ refer to the untrapped and antitrapped magnetic
sublevels, respectively. Here we included the gravitational potential and
denote by $V(0)$ the offset of the trapping potential for the $m=-1$
condensate in the geometrical center of the trap at $r=0$. For simplicity we
shall assume that the magnetic trapping potential is isotropic. The
$rf$-frequency is denoted by $\omega$ and $\Omega=g\mu_B|B_{rf}|/\sqrt{2}\hbar$
is the Rabi frequency at resonance due to the magnetic $rf$-field, which
is spatially independent over the condensate.

Similarly, for $rf$-outcoupling from an $F=2$, $m_F=+2$ condensate (cf. fig.\ref{1}(b)), like the
$^{87}Rb$ condensate, $m=-2,-1,0,1,2$ labels the 5 magnetic
sublevels of the $F=2$ state with trapping potentials
\begin{eqnarray}
V_{\pm2} &=& \pm\left(V(0)+\frac{M}{2}\bar{\omega}^2r^2\right)+Mg(z-z_0)
 \nonumber \\
V_{\pm1} &=& \pm\frac{1}{2}
  \left(V(0)+\frac{M}{2}\bar{\omega}^2r^2\right)+Mg(z-z_0)\\
\label{eq:2.3}
V_0 &=& Mg(z-z_0)\nonumber
\end{eqnarray}
The coupling term between magnetic sublevels in eq.(\ref{eq:2.1}) now reads
\cite{F1}, \cite{F2}
\[
\frac{1}{2}\hbar\Omega (\sqrt{(2-m)(2+m+1)}e^{i\omega t}\psi_{m+1}+\sqrt{(2+m)(2-m+1)}e^{-i\omega t}\psi_{m-1})
\]
with $\Omega=g\mu_B|B_{rf}|/\hbar$.

For microwave outcoupling between two different hyperfine states
it is sufficient, due to energy conservation, to consider only two components
$m=1$, $m=2$ in eq.~(\ref{eq:2.1}) which refer to the trapped and untrapped
(or antitrapped) states, respectively. With the magnetic dipole matrix-element
$\mu_{12}$ between the two states , the Rabi frequency is $\Omega=\mu_{12}|B|/2\hbar$ where the magnetic microwave field is again spatially
independent across the condensate. In this case the potentials $V_m$ are
\begin{eqnarray}
V_1 &=&
\frac{M}{2}\bar{\omega}^2r^2+V(0)+Mg(z-z_0)\nonumber \\
V_2 &=& -p\left(\frac{M}{2}\bar{\omega}^2r^2+V(0)\right)+Mg(z-z_0)
\label{eq:2.4}
\end{eqnarray}
where $p=0$ and $p>0$ for outcoupling to an untrapped state and an antitrapped
state, respectively.

For Raman outcoupling (cf. fig.\ref{1}(c)) $m=1$, $m=2$ again refer to the trapped
and untrapped (or antitrapped) state respectively, and $\Omega$ is replaced
by $\Omega\cdot e^{i\bbox{k}\bbox{x}}$, the space-dependent effective Rabi
frequency of the Raman transition, with net energy and momentum transfer to
the atom of $\hbar\omega$ and $\hbar\bbox{k}$, respectively. In a three-level
approximation for the Raman transition
$\Omega=\frac{\Omega_1\Omega_2}{\delta}$. Here $\Omega_1$, $\Omega_2$ are
the Rabi frequencies of the two lasers with frequencies $\omega_1$,
$\omega_2$ driving transitions between levels 1 and 3, and levels 2 and 3,
respectively, with $\omega=\omega_1-\omega_2$ and $\bbox{k}=\bbox{k}_1-
\bbox{k}_2$. $\Omega$ depends on the detuning $\delta=(E_3-E_1)/\hbar-\omega_1$
between the level-spacing $E_3-E_1$ and $\hbar\omega_1$, which is assumed to
be large to avoid the population of the auxiliary level 3. The potentials
$V_m$ in this case are as in eqs.~(\ref{eq:2.4}).

In the following we shall assume that the electromagnetic outcoupling fields
are switched on at time $t=0$. We shall consider the case of sudden switch-on
in a time interval short compared to the final Rabi period. Before the
switch-on the condensate state is assumed to be in equilibrium and described
to sufficient accuracy by the Thomas-Fermi approximation, while the
outcoupled states are unpopulated.

\section{Rabi-oscillations in the condensate}\label{sec:3}

\begin{figure}
\centerline{
\epsfig{figure=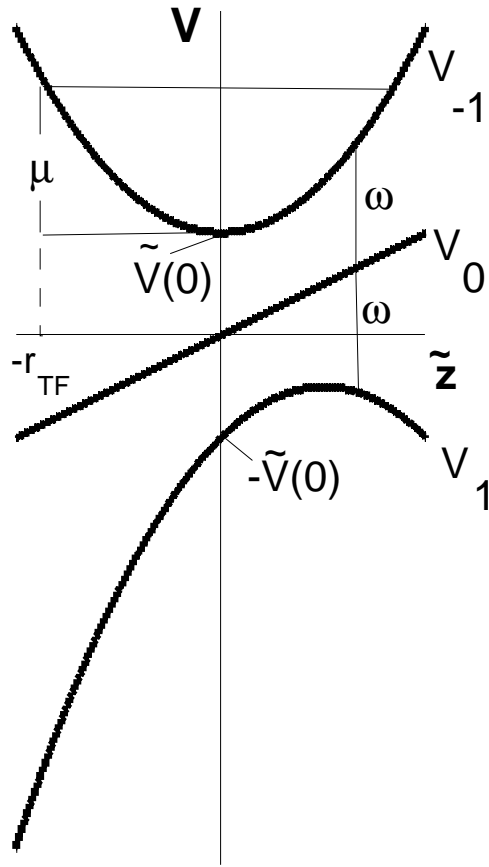}}
\caption{
Potentials of the Zeeman sublevels of the $F=1$ manifold \label{2}
}
\end{figure}

For sufficiently strong external electromagnetic outcoupling fields the Rabi
oscillations  between the coupled internal atomic states become so fast, that
the center of mass motion of the atoms can no longer follow. In
this regime a combined Thomas-Fermi and Raman-Nath approximation
applies, in which the kinetic energy terms in eq.~(\ref{eq:2.1}) are
negligible. The Rabi oscillations within a given $F$ hyperfine manifold
can then be solved analytically. We consider the cases $F=1$ and $F=2$ in turn:

{\bf 1. Rabi oscillations in an $F=1$ manifold}:\\
Before starting let us shift the origin of the $z$-axis into the minimum of the
total external potential (trap and gravity) seen by the condensate in the $m=-1$ state via the
replacement
\[
z\to\tilde{z}=z+g/\bar{\omega}^2\,.
\]
Furthermore let us choose the zero of the gravitational potential in
$\tilde{z}=0$ by adopting $z_0=-g/\bar{\omega}^2$. With the notation
\begin{equation}
\tilde{r}^2=x^2+y^2+\tilde{z}^2\,,\quad
\tilde{V}(0)=V(\tilde{r}=0)=V(0)+Mg^2/2\bar{\omega}^2
\label{eq:3.1}
\end{equation}
we then have
\begin{eqnarray}
V_{-1} &=& \frac{M}{2}\bar{\omega}^2\tilde{r}^2+\tilde{V}(0)\nonumber\\
V_0 &=& Mg\tilde{z}\\
\label{eq:3.2}
V_{1}&=& -\frac{M}{2}\bar{\omega}^2\tilde{r}^2-\tilde{V}(0)+2Mg\tilde{z}
\nonumber
\end{eqnarray}
(see fig.\ref{2}).\\
Furthermore let us proceed to an interaction picture by splitting off the main
frequencies according to
\begin{eqnarray}
\psi_{-1}&=&
e^{-i(\mu+\tilde{V}(0))t/\hbar}\tilde{\psi}_{-1}\nonumber\\
\psi_0 &=&
e^{-i(\mu+\tilde{V}(0))t/\hbar}e^{i\omega t}\tilde{\psi}_0\\
\label{eq:3.3}
\psi_{1} &=&
e^{-i(\mu+\tilde{V}(0))t/\hbar}e^{2i\omega t}\tilde{\psi}_1\,.\nonumber
\end{eqnarray}

Initial conditions have to be specified next. We shall assume the
electromagnetic outcoupling fields are switched on at $t=0$ in a time interval
short compared to the final Rabi period. The initial condition at $t=0$ in the Thomas-Fermi approximation \cite{TF1}, \cite{TF2} then
is
\begin{eqnarray}
\psi_0 &=& 0 = \psi_1\\
\label{eq:3.4}
\psi_{-1}&=& e^{-\frac{i}{\hbar}(\mu+\tilde{V}(0))t}
 \sqrt{\frac{\mu-\frac{M}{2}\bar{\omega}^2\tilde{r}^2}{U_0}}
  \Theta (\mu-\frac{M}{2}\bar{\omega}^2\tilde{r}^2)\nonumber
\end{eqnarray}
where the chemical potential $\mu$ is determined by the number of atoms in the condensate via
\begin{equation}
N=\frac{4\pi}{15U_0}(2\mu)^{5/2}/(M\bar{\omega}^2)^{3/2}\,.
\end{equation}

\begin{figure}
\centerline{
\epsfig{figure=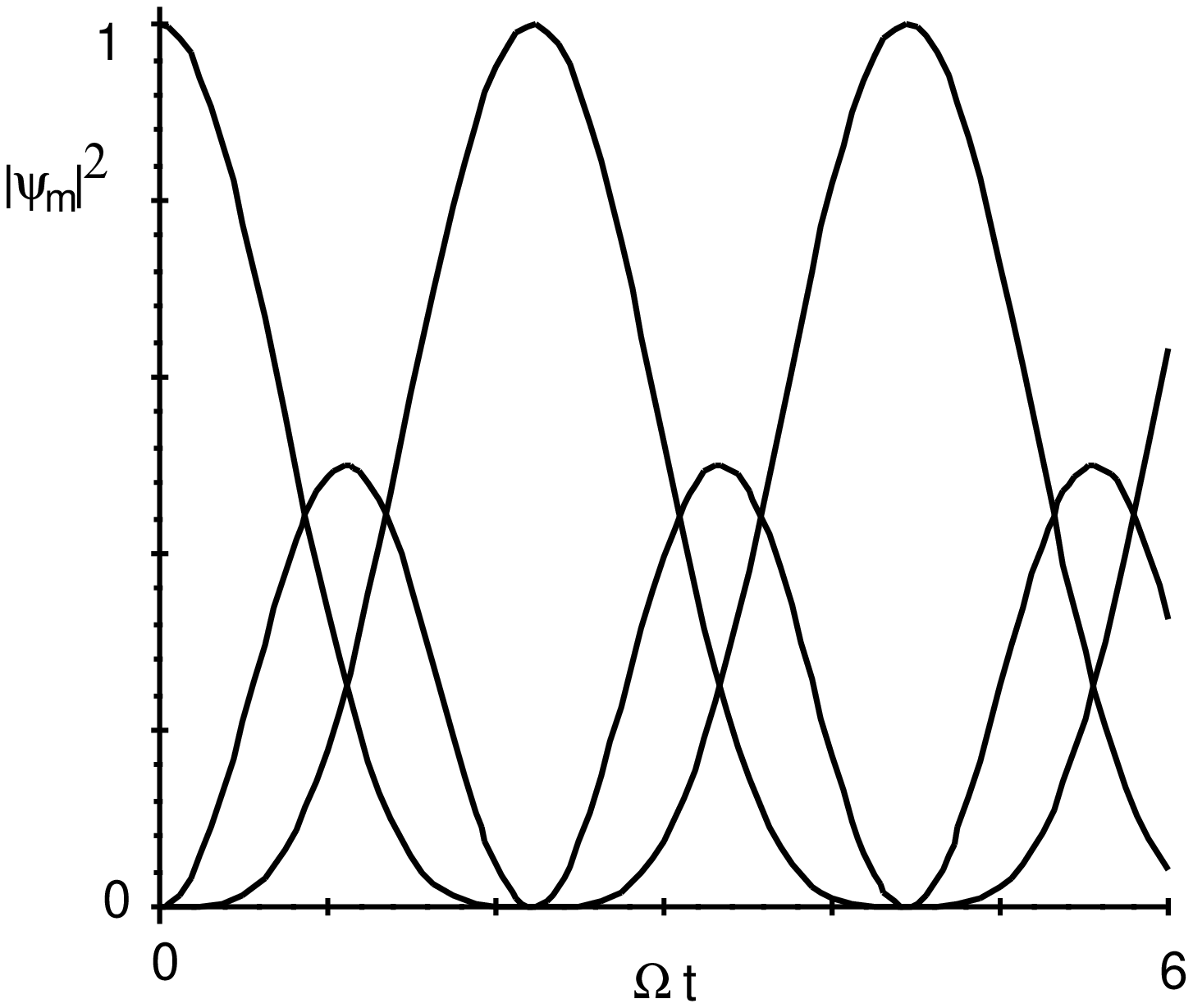}}
\caption{
Occupation probabilities of the Zeeman sublevels of the $F=1$ manifold \label{3}
}
\end{figure}

Neglecting the kinetic energy terms in a zeroth-order approximation, the coupledGross-Pitaevskii equations reduce to a Rabi level-flopping problem, in which
the local quantity $|\psi(\bbox{\tilde{r}},t)|^2$ is constant in time. In the next
(first-order) approximation $|\psi(\bbox{\tilde{r}},t)|^2$ will be allowed to
change slowly in space and time and will then be determined self-consistently. Therefore, in the present zeroth order approximation
we have to treat $|\psi(\bbox{\tilde{r}},t)|^2$ as constant in time, but
arbitrary. In a vector and matrix notation, with
\[
\tilde{\psi}=
\left(\begin{array}{c}
\tilde{\psi}_{-1}\\
\tilde{\psi}_0\\
\tilde{\psi}_1\end{array}\right)
\]
the coupled Gross-Pitaevskii equations to zeroth order then reduce to the simple form
\begin{equation}
i\dot{\tilde{\psi}}=
\left(\begin{array}{lcc}
\varepsilon & \Omega & 0\\
\Omega & \varepsilon+\Delta & \Omega\\
0 & \Omega & \varepsilon+2\Delta\end{array}\right)\tilde{\psi}
\label{eq:Os}
\end{equation}
with
\begin{eqnarray}
\hbar\varepsilon(\bbox{\tilde{r}},t)&=&
U_0|\tilde{\psi}(\bbox{\tilde{r}},t)|^2-\mu+\frac{M}{2}
\bar{\omega}^2\tilde{r}^2\nonumber\\
\hbar\Delta(\bbox{\tilde{r}}) &=& -\frac{M}{2}\bar{\omega}^2\tilde{r}^2+
 Mg\tilde{z}-\tilde{V}(0)+\hbar\omega
\label{eq:3.6a}
\end{eqnarray}
Its solution is recorded in the appendix, together with the corresponding solutions for the other cases considered below,
in particular the case $F=2 $.
Let us define the regime of strong outcoupling by the condition
\begin{equation}
\Omega\gg|\Delta(\bbox{\tilde{r}})|
\label{eq:sc}
\end{equation}
throughout the condensate. If the $rf$-frequency $\omega$ is chosen in such
a way that the resonance condition $\Delta (\bbox{\tilde{r}})=0$ is satisfied
somewhere inside the condensate, a sufficient condition for (\ref{eq:sc}) to
hold is $\Omega\gg\mu$.
Since in the regime of strong outcoupling the detuning $\Delta $ is negligible
compared to $\Omega$, the solution of eq.~(\ref{eq:Os}) becomes very
simple and reads
\begin{equation}
\tilde{\psi}(\bbox{\tilde{r}},t)=
\frac{1}{2}\sqrt{|\psi(\bbox{\tilde{r}},t)|^2}
e^{-i\int_0^td\tau\,[\varepsilon(\bbox{\tilde{r}},\tau)+
   \Delta(\bbox{\tilde{r}})]}\cdot
   \left(\begin{array}{l}
   1+\cos\sqrt{2}\Omega t\\
   -i\sqrt{2}\sin\sqrt{2}\Omega t\\
   -1+\cos\sqrt{2}\Omega t
   \end{array}\right)\,.
\label{eq:3.7b}
\end{equation}
Fig.\ref{3} shows the probabilities $|\tilde{\psi}_m(\bbox{\tilde{r}},t)|^2/|\psi(\bbox{\tilde{r}},t)|^2$ as
a function of time. Thus the Rabi oscillations in the strong outcoupling regime occur with
approximately constant amplitude throughout the condensate, but with a
space-dependent phase. In the next order approximation the gradient of this
phase gives rise to a particle current, which in turn changes the number
density $|\psi(\bbox{\tilde{r}},t)|^2$ by particle number conservation. This
will be considered in section \ref{sec:4}.

\begin{figure}\centerline{
\epsfig{figure=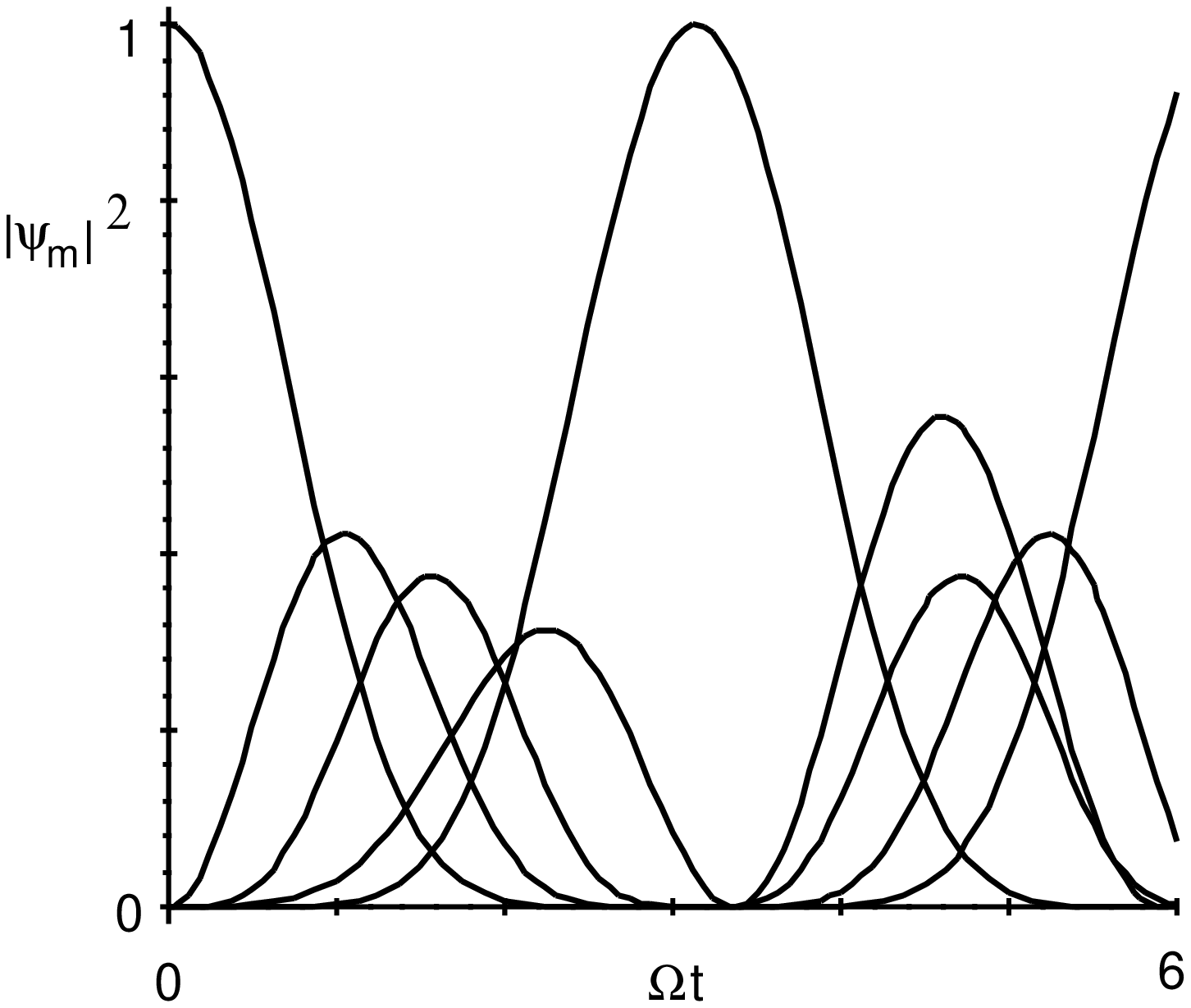}}
\caption{
Occupation probabilities of the Zeeman sublevels of the $F=2$ manifold \label{4}
}
\end{figure}

{\bf 2. Rabi oscillations in an $F=2$ manifold}\\
Let us assume the condensate is in the $F=2$, $m_F=2$ state. We perform an
analogous sequence of steps as in the previous subsection for the case
$F=1$ Introducing $\tilde{z}=z+g/\bar{\omega}^2$ and choosing
$z_0=-g/\bar{\omega}^2$ we have in the notation (\ref{eq:3.1})
\begin{eqnarray}
V_2 &=& \frac{M}{2}\bar{\omega}^2\tilde{r}^2+\tilde{V}(0)\nonumber\\
V_1 &=& \frac{1}{2}
 \left(\frac{M}{2}\bar{\omega}^2\tilde{r}^2+\tilde{V}(0)\right)+
  \frac{1}{2}Mg\tilde{z}\nonumber\\
V_0 &=& Mg\tilde{z}\label{eq:3.7c}\\
V_{-1} &=& -\frac{1}{2}
 \left(\frac{M}{2}\bar{\omega}^2\tilde{r}^2+\tilde{V}(0)\right)+
  \frac{3}{2}Mg\tilde{z}\nonumber\\
V_{-2}&=& -\left(\frac{M}{2}\bar{\omega}^2\tilde{r}^2+\tilde{V}(0)\right)+
  2Mg\tilde{z}\nonumber\,.
\end{eqnarray}
Let us split off the main frequencies according to
\begin{equation}
\psi_m=e^{-i(\mu+\tilde{V}(0))t/\hbar}
 e^{-i(m-2)\omega t}\tilde{\psi}_m\,.
\label{eq:3.7d}
\end{equation}
Using the fact that $|\psi(t)|^2=\sum_m|\psi_m(t)|^2$ is locally conserved in
the combined Thomas-Fermi and Raman-Nath approximation and left arbitrary at
this stage (to be determined self-consistently later on) we obtain the coupled
set of equations, in matrix notation with
\[
\tilde{\psi}=\left(\begin{array}{c}
 \tilde{\psi}_2\\
 \tilde{\psi}_1\\
 \vdots\\
 \tilde{\psi}_{-2}
\end{array}\right)
\]
\begin{equation}
i\dot{\tilde{\psi}}=
\left(\begin{array}{ccccc}
\varepsilon &\Omega & 0 & 0 & 0\\
\Omega & \varepsilon+\Delta & \sqrt{\frac{3}{2}}\Omega & 0 & 0\\
0 & \sqrt{\frac{3}{2}}\Omega & \varepsilon+2\Delta & \sqrt{\frac{3}{2}}
\Omega & 0\\
0 & 0 & \sqrt{\frac{3}{2}}\Omega & \varepsilon+3\Delta & \Omega\\
0 & 0 & 0 & \Omega & \varepsilon+4\Delta\end{array}\right)\tilde{\psi}
\label{eq:F21}
\end{equation}
where
\begin{eqnarray}
 \hbar\varepsilon &=& U_0|\psi|^2-\mu+\frac{M}{2}\bar{\omega}^2\tilde{r}^2
\nonumber\\
 \hbar\Delta &=& -\frac{1}{2}\left(\frac{M}{2}\bar{\omega}^2\tilde{r}^2+
  \tilde{V}(0)\right)+\frac{1}{2}Mg\tilde{z}+\hbar\omega\,.
\label{eq:F22}
\end{eqnarray}
The solution of (\ref{eq:F21}) is given in the appendix. In the limit of strong outcoupling it takes the form
\begin{equation}
\tilde{\psi}=
\frac{1}{2}\sqrt{|\psi|^2}e^{-i\int_0^td\tau\,
 [\varepsilon(\bbox{\tilde{r}},\tau)+2\Delta(\bbox{\tilde{r}})]}
 \cdot
 \left(\begin{array}{l}
 \cos\Omega t+\frac{1}{4}\cos 2\Omega t+\frac{3}{4}\\
  -i\sin\Omega t-\frac{i}{2}\sin 2\Omega t\\
  \frac{\sqrt{6}}{4}\left(\cos 2\Omega t-1\right)\\
  i\sin\Omega t-\frac{i}{2}\sin 2\Omega t\\
  -\cos\Omega t+\frac{1}{4}\cos 2\Omega t+\frac{3}{4}
  \end{array}\right)\,.
\label{eq:3.10}
\end{equation}
Fig.\ref{4} shows the probabilities $|\tilde{\psi}_m(\bbox{\tilde{r}},t)|^2/|\psi(\bbox{\tilde{r}},t)|^2$ as
a function of time for this case.

{\bf 3. Rabi oscillations for microwave and Raman outcoupling}\\
For these outcoupling mechanisms the internal state of the condensate atoms
is coupled to a single magnetic sublevel in another hyperfine manifold and
a simple two-level Rabi oscillation results. As before we introduce
$\tilde{z}=z+g/\bar{\omega}^2$ and pick $z_0=-g/\bar{\omega}^2$. Then
\begin{eqnarray}
V_1 &=& \frac{M}{2}\bar{\omega}^2\tilde{r}^2+\tilde{V}(0)\nonumber\\
V_2 &=& -p\left(\frac{M}{2}\bar{\omega}^2\tilde{r}^2+\tilde{V}(0)\right)+
 Mg(1+p)\tilde{z}\,.
\label{eq:3.11}
\end{eqnarray}
We define $\tilde{\psi}_m$ via
\begin{eqnarray}
\psi_1 &=& e^{-i(\mu+\tilde{V}(0))t/\hbar}\tilde{\psi}_1\nonumber\\
\psi_2 &=& e^{-i(\mu+\tilde{V}(0))t/\hbar+i\omega t}\tilde{\psi}_2
 e^{ik\tilde{z}}
\label{eq:3.12}
\end{eqnarray}	
where $k\ne0$ only in the case of Raman outcoupling. Then the equations in
combined Thomas-Fermi and Raman-Nath approximation in a matrix representation
with $\tilde{\psi}={\tilde{\psi}_1\choose\tilde{\psi}_2}$ read
\begin{equation}
 i\dot{\tilde{\psi}} =
 \left(\begin{array}{ll}
 \varepsilon & \Omega\\
 \Omega & \varepsilon+\Delta\end{array}\right)\tilde{\psi}
\label{eq:3.13}
\end{equation}
with
\begin{eqnarray}
\hbar\varepsilon &=& U_0|\psi|^2-\mu+\frac{M}{2}\bar{\omega}^2\tilde{r}^2\nonumber\\
\hbar\Delta &=& \frac{\hbar^2 k^2}{2M}-(1+p)
 \left(\frac{M}{2}\bar{\omega}^2\tilde{r}^2-Mg\tilde{z}+\tilde{V}(0)\right)+
  \hbar\omega\,.
\label{eq:3.14}
\end{eqnarray}
Their solution in the limit of strong outcoupling $\Omega \gg\Delta$ is
\begin{equation}
\tilde{\psi}(t) =
\sqrt{(\psi(0)|^2}\left(\begin{array}{l}
\cos\Omega t\\
-i\sin\Omega t\end{array}\right)e^{-i\int_0^td\tau\,
 \left(\varepsilon(\bbox{\tilde{r}},\tau )+\frac{1}{2}\Delta(\bbox{\tilde{r}})
  \right)}\,.
\label{eq:3:15}
\end{equation}

\section{Current of atoms with outcoupling fields on}\label{sec:4}

In the preceding section the kinetic energy term in the Gross-Pitaevskii
equation (\ref{eq:2.1}) was neglected. No spatial transport of particles
can occur in this approximation, the atoms are assumed to be at rest. The most
important consequence of the kinetic energy term in eq.~(\ref{eq:2.1}) is that
a spatial transport of the atoms occurs. It obeys the conservation law
\begin{equation}
\frac{\partial|\psi|^2}{\partial t}+\bbox{\nabla}\cdot\bbox{j}=0
\label{eq:4.1}
\end{equation}
with the current density
\begin{equation}
\bbox{j}=\frac{\hbar}{2Mi}\sum_m
\big(\psi_m^*\bbox{\nabla}\psi_m-\psi_m\bbox{\nabla}\psi_m^*\big)
\label{eq:4.2}
\end{equation}
We shall now calculate this current using the results obtained in the
preceding section.

The present discussion applies to all cases 1) to 3) of sec. \ref{sec:3}. Let
us begin with the case of $rf$-outcoupling
from a $F=1$, $m_F=-1$ condensate. The state-vector $\tilde{\psi}$ is
space-dependent due to its proportionality to $\sqrt{|\psi(\tilde{r},t)|^2}$
and its dependence on $\varepsilon$ and $\Delta$ which both depend on
$\tilde{\bbox{r}}$. As shown in the appendix
$\tilde{\psi}$ can be written as
\begin{equation}
\tilde{\psi}=|\psi|e^{i\tilde{\phi}}\varphi
\label{eq:4.4}
\end{equation}
where the vector $\varphi$ is space-dependent only via its dependence on the space-dependent detuning $\Delta (\bbox{\tilde{r}})$ and has unit norm. The phase
 $\tilde{\phi}$, according to eq.~(\ref{eq:3.7b}), satisfies in the local rest frame of
the atoms
\begin{equation}
\frac{\partial\tilde{\phi}}{\partial t}=-\left(\varepsilon(\bbox{\tilde{r}}, t)+
 \Delta(\bbox{\tilde{r}})\right)\,.
\label{eq:4.5}
\end{equation}
Inserting (\ref{eq:4.4}) in the definition of the current density results in the
expression
\begin{equation}
\bbox{j}(\bbox{\tilde{r}},t)=\bbox{v}(\bbox{\tilde{r}},t)
 |\psi(\bbox{\tilde{r}},t)|^2
\label{eq:4.6}
\end{equation}
with the local velocity field in the laboratory frame
\begin{equation}
 \bbox{v}(\bbox{\tilde{r}},t)=\frac{\hbar}{M}\bbox{\nabla}\tilde{\phi}+\frac{\hbar}{2Mi}\sum_m
\big(\varphi_m^*\bbox{\nabla}\varphi_m-\varphi_m\bbox{\nabla}\varphi_m^*\big)
\,.
\label{eq:4.7}
\end{equation}
As is shown in the appendix the second term on the right hand side of eq.~(\ref{eq:4.7}) is given by
\begin{equation}
-\frac{\sqrt{2}\bar{\omega}}{2\Omega}sin(\sqrt{2}\Omega t)(\bar{\omega}\tilde{\bbox{r}}-\frac{g}{\bar{\omega}}\bbox{e}_z)
\label{eq:4.7a}
\end{equation}
For strong outcoupling it is rapidly oscillating on the time-scale $\bar{\omega}^{-1}$ of the center of mass motion and can therefore be neglected in the following. If taken into account it would describe a small oscillatory component in the temporal evolution of the center of mass and the radius of the condensate.

The equation of motion of $\tilde{\phi}$ in the laboratory frame includes the kinetic
energy $M\bbox{v}^2/2=\hbar^2(\bbox{\nabla}\tilde{\phi})^2/2M$ in addition to $\Delta$ and $\epsilon$ and reads
\begin{equation}
\frac{\partial\tilde{\phi}}{\partial t}=
-\left(\hbar(\bbox{\nabla}\tilde{\phi})^2/2M+\varepsilon(\bbox{\tilde{r}},t)+
\Delta(\bbox{\tilde{r}})\right)\,.
\label{TE}
\end{equation}
It has to be solved together with the continuity equation
\begin{equation}
\frac{\partial|\psi|^2}{\partial t}+\frac{\hbar}{M}\bbox{\nabla}\cdot
(|\psi|^2\bbox{\nabla}\tilde{\phi})=0\,.
\label{eq:TE1}
\end{equation}
Before doing this it is convenient to make a transformation to the rest-frame
of the center of mass
\begin{equation}
\tilde{\psi}(\bbox{\tilde{r}},t) =
e^{-\frac{i}{\hbar}\,\frac{M}{2}\int_0^t\dot{\bbox{\tilde{r}}}^{*2}(\tau)d\tau+
 \frac{iM}{\hbar}\dot{\bbox{\tilde{r}}}^*(t)\cdot\bbox{\tilde{r}}}
  \tilde{\psi}_x(\bbox{x},t)
\label{eq:4.10}
\end{equation}
with $\tilde{\psi}_x=|\psi_x|e^{i\tilde{\phi}_x}$ and
\begin{equation}
\bbox{x}=\bbox{\tilde{r}}-\dot{\bbox{\tilde{r}}}^*(t)\,.
\label{eq:4.11}
\end{equation}
The continuity and phase equations in the center of mass frame read
\begin{equation}
\frac{\partial|\psi_x(\bbox{x},t)|^2}{\partial t}+
\bbox{\nabla}\cdot\bbox{u}|\psi_x(\bbox{x},t)|^2=0
\label{eq:4.12}
\end{equation}
with
\begin{equation}
\bbox{u}=\frac{\hbar}{M}\bbox{\nabla}\tilde{\phi}_x(\bbox{x},t)
\label{eq:4.13}
\end{equation}
and
\begin{equation}
\frac{\partial\tilde{\phi}_x}{\partial t}
=
-\frac{\hbar}{2M}(\bbox{\nabla}\tilde{\phi}_x)^2+
 M\ddot{\bbox{\tilde{r}}}^*(t)\cdot\bbox{x}-
  \varepsilon\left(\bbox{\tilde{r}}^*(t)+\bbox{x},t\right)-
  \Delta\left(\bbox{\tilde{r}}^*(t)+\bbox{x}\right)\,.
\label{eq:4.14}
\end{equation}
With the center of mass motion $\bbox{\tilde{r}}^*=
-\frac{1}{2}gt^2\bbox{e}_z$ and eqs.(\ref{eq:3.6a}) the phase equation simplifies to
\begin{equation}
\frac{\partial\tilde{\phi}_x}{\partial t}
=
-\frac{\hbar}{2M}(\bbox{\nabla}\tilde{\phi}_x)^2-\frac{U_0}{\hbar}|\psi_x|^2+
 \frac{\mu+\tilde{V}(0)}{\hbar}-\omega-\frac{Mg}{\hbar}\tilde{z}^*\,.
\label{eq:4.15}
\end{equation}
The ansatz
\begin{eqnarray}
|\psi_x|^2 &=&
 \left(A(t)-B(t)x^2\right)\Theta\left(A(t)-B(t)x^2\right)\nonumber\\
\tilde{\phi}_x &=&
 \alpha(t)-\beta(t)x^2
\label{eq:A}
\end{eqnarray}
solves equations (\ref{eq:4.12}), (\ref{eq:4.13}), (\ref{eq:4.15})
 exactly
if the coefficients satisfy the differential equations
\begin{eqnarray}
\dot{B} &=&
-\frac{10\hbar}{M}\beta B\nonumber\\
\dot{\beta} &=&
\frac{U_0}{\hbar}B-\frac{2\hbar}{M}\beta^2\\
\label{eq:4.17}
\dot{\alpha} &=&
-\frac{U_0}{\hbar}A\nonumber
\end{eqnarray}
and, by the normalization condition $N=\int d^3x |\psi_x|^2$,
\begin{equation}
A=\left(\frac{15N}{8\pi}\right)^{2/5}B^{3/5}\,.
\label{eq:4.18}
\end{equation}
 The initial
conditions follow from the Thomas-Fermi wave function at $t=0$. They are
\begin{equation}
B(0) = \frac{M\bar{\omega}^2}{2U_0}\,,\quad \alpha(0)=0=\beta(0)\,.
\label{eq:4.19}
\end{equation}
The time-dependent Thomas-Fermi radius is related to $B(t)$ via
\begin{equation}
r_{TF}(t)=r_{TF}(0)
 \left(\frac{B(0)}{B(t)}\right)^{1/5}\,.
\label{eq:4.20}
\end{equation}
Its equation of motion follows as
\begin{equation}
\ddot{r}_{TF}(t)-\frac{\bar{\omega}^2r_{TF}^5(0)}{r_{TF}^4(t)}=0\,.
\label{eq:4.21a}
\end{equation}
It can be integrated once to give
\begin{equation}
\dot{r}_{TF}^2(t)+\frac{2\bar{\omega}^2}{3}\,
 \frac{r_{\tiny{TF}}^5(0)}{r_{\tiny{TF}}^3(t)}=
 \frac{2}{3}\bar{\omega}^2r_{TF}^2(0)\,.
\label{eq:4.22}
\end{equation}
At large times, where $r_{TF}(t)\gg r_{TF}(0)$, the Thomas-Fermi radius
expands ballistically at a constant rate
\begin{equation}
\dot{r}_{TF}(\infty)=\sqrt{\frac{2}{3}}\bar{\omega}r_{TF}(0).
\label{eq:4.23}
\end{equation}
At small times it expands with acceleration $\bar{\omega}^2r_{TF}(0)$
\begin{equation}
 r_{TF}(t)\simeq r_{TF}(0)\left(1+\frac{1}{2}\bar{\omega}^2t^2\right)\,.
\label{eq:4.24}
\end{equation}
The Thomas-Fermi radius  obtained by integrating
eq.(\ref{eq:4.22}) is plotted in fig.\ref{5} as a function of time.

The result for the condensate density in the laboratory frame is
\begin{equation}
|\psi(t)|^2=\frac{15N}{8\pi}\,\frac{r_{TF}^2(t)-
 \left(\bbox{\tilde{r}}+\frac{1}{2}gt^2\bbox{e}_z\right)^2}{r_{TF}^5(t)}\,
  \Theta\left(r_{TF}^2(t)-
   \left(\bbox{\tilde{r}}+\frac{1}{2}gt^2\bbox{e}_z\right)^2\right)\,.
\label{eq:4.25}
\end{equation}
The phase of the macroscopic state vector $\tilde{\psi}$ can also be expressed
by $r_{TF}(t)$ as
\begin{equation}
\tilde{\phi}(t) = \tilde{V}(0)\frac{t}{\hbar}-\omega t-\frac{Mg}{\hbar}t\tilde{z}-
 \frac{\mu}{\hbar}\int_0^td\tau
  \left(\frac{r_{TF}^3(0)}{r_{TF}^3(\tau)}-1\right)d\tau+\frac{M}{2\hbar}
  \left(\bbox{\tilde{r}}-\bbox{\tilde{r}}^*(t)\right)^2\frac{d}{dt}\ln
   \frac{r_{TF}(t)}{r_{TF}(0)}
\label{eq:4.26}
\end{equation}
Thus under the influence of the strong electromagnetic outcoupling fields,
switched on at $t=0$, the whole condensate undergoes rapid Rabi oscillations
through all the coupled states while freely dropping out of the trap by
gravity, and at the same time expanding due to the repulsive interaction.
Switching on the external electromagnetic field therefore effectively switches
off the trapping potential, up to the oscillatory term (\ref{eq:4.7a}) whose influence is small if $\bar{\omega}<<\Omega$, because it averages out in the Rabi oscillations
through trapped, untrapped and antitrapped states of a hyperfine manifold.

The results given so far have been written down for $rf$-outcoupling from an
$F=1$, $m_F=-1$ condensate (\ref{sec:3} 1), but are easily transferred to the
other cases considered in section (\ref{sec:3}):\\

{\bf RF-Outcoupling from the $F=2$, $m_F=2$ state:}\\
 The velocity field $\bbox{v}$ is still given by eq.(\ref{eq:4.7}), but the second term on the right hand side of that equation is changed and becomes now
 \begin{equation}
 -\frac{\bar{\omega}}{\Omega}sin (\Omega t)(\bar{\omega}\tilde{\bbox{r}}-\frac{g}{\bar{\omega}}\bbox{e}_z)\,.
\label{eq:4.26a}
      \end{equation}
 It is again rapidly oscillating on the relevant time-scale $\bar{\omega}$
and hence negligible.
    The equation of motion of the phase becomes
      \begin{equation}
      \frac{\partial\tilde{\phi}}{\partial t}=-
 \left(\frac{\hbar}{2M}(\bbox{\nabla}\tilde{\phi})^2+\varepsilon+2\Delta\right)\,.\label{eq:4.27}
      \end{equation}
      After inserting $\varepsilon$ and $\Delta$ from eq.~(\ref{eq:F22}),
      the phase equation reduces to the same form as in the case
      of the $F=1$ manifold and therefore the results for $\bbox{\tilde{r}}^*(t), r_{TF}(t),|\psi|^2$ and
      $\tilde{\phi}$ are unchanged.

{\bf Microwave and Raman outcoupling:}\\
       We give the results for Raman outcoupling. To obtain the results for
       microwave outcoupling from the following formulae on has to set $k=0$
       there. The velocity is again given by the general expression (\ref{eq:4.7}), with the second term now of the form
\[
\frac{(1+p)\bar{\omega}}{4\Omega}sin(2\Omega t)(\bar{\omega}\tilde{\bbox{r}}-\frac{g}{\bar{\omega}}\bbox{e}_z),
\]
 which is again rapidly oscillating on the time-scale $\bar{\omega}$ and therefore negligible.
The phase equation then takes the form
       \begin{equation}
       \frac{\partial\tilde{\phi}}{\partial t}=-
       \left(
     \frac{\hbar}{2M}(\bbox{\nabla}\tilde{\phi})^2+\varepsilon+\frac{\Delta}{2}
       \right)
       \label{eq:4.28}
       \end{equation}
       with
       \begin{eqnarray}
       \hbar\left(\varepsilon+\frac{\Delta}{2}\right)= && \hbar\omega/2
       +\hbar
       U_0|\psi|^2-\mu+\frac{\hbar^2k^2}{4M}+\frac{1}{2}(1-p)\frac{M}{2}
       \bar{\omega}^2\tilde{r}^2\nonumber \\
       &&\qquad +\frac{1}{2}(1+p)Mg\tilde{z}-\frac{1}{2}
       (1+p)\tilde{V}(0)\,.
       \label{eq:4.29}
       \end{eqnarray}

Let us first consider the case $p=1$. There the excited state is antitrapped and has a magnetic
moment precisely opposite to the trapped condensate state. The trap potential thus averages out in the Rabi
oscillations. We transform to the center of mass frame with
\begin{equation}
 \dot{\bbox{\tilde{r}}}^*=\frac{\hbar\bbox{k}}{M}-gt\bbox{e}_z\,.
\label{eq:4.28a}
\end{equation}
There we can again make the ansatz (\ref{eq:A}) and solve for $A(t)$, $B(t)$
as before. Finally, transforming back to laboratory coordinates we obtain
\begin{equation}
|\psi(t)|^2= \frac{15N}{8\pi}\,
\frac{r_{TF}^2(t)-\left(\bbox{\tilde{r}}+
 \frac{1}{2}\bbox{e}_zgt^2-\frac{\hbar \bbox{k}}{M}t\right)^2}{r_{TF}^5(t)}
\label{eq:4.30}
\end{equation}
with the previous result (\ref{eq:4.21a}, \ref{eq:4.22}) for the equation of motion of $r_{TF}(t)$.

In the case $p\ne1$,
to which we now turn, the trap potential does not average out in the Rabi
oscillations and contributes to the center of mass motion and the spreading
of the condensate. Now the transformation (written for $p<1$)
$\bbox{\tilde{r}}=\bbox{\tilde{r}}^*(t)+\bbox{x}$ with
\begin{equation}
\bbox{\tilde{r}}^*(t) = -\frac{1+p}{1-p}\,\frac{g\bbox{e}_z}{\bar{\omega}^2}
 \left(1-\cos\left(\sqrt{\frac{1-p}{2}}\bar{\omega}t\right)\right)
  +\frac{\hbar\bbox{k}}{M}\,
  \frac{\sin\left(\sqrt{\frac{1-p}{2}}\bar{\omega}t\right)}
        {\sqrt{\frac{1-p}{2}}\bar{\omega}}
\label{eq:4.31}
\end{equation}
introduces the center of mass rest-frame. The phase equation in this frame takes the form
\begin{eqnarray}
\frac{\partial\tilde{\phi}_x}{\partial t}&=&
-\frac{\hbar}{2M}(\bbox{\nabla}\tilde{\phi}_x)^2-\frac{U_0}{\hbar}|\psi_x|^2-
 \frac{1-p}{2\hbar}\,\frac{M}{2}\bar{\omega}^2x^2\nonumber\\
&& \\
\label{eq:4.32}
&&-\frac{1-p}{2\hbar}\,\frac{M}{2}\bar{\omega}^2\tilde{r}^{*2}-
    \frac{1+p}{2\hbar}Mg\tilde{z}^*+\frac{\mu}{\hbar}-
     \frac{\hbar k^2}{4M}+\frac{1+p}{2\hbar}\tilde{V}(0)-\frac{\omega}{2}
\nonumber
\end{eqnarray}
Again the continuity equation and the phase equation can be solved by an
ansatz of the form (\ref{eq:A}).
The solution for the condensate number density with the required initial
conditions takes the form, in the laboratory frame,
\begin{equation}
|\psi(t)|^2 = \frac{15N}{8\pi}\,
 \frac{r_{TF}^2(t)-
  \left(\bbox{\tilde{r}}-\bbox{\tilde{r}}^*(t)\right)^2}{r_{TF}^5(t)}\,
   \Theta\left(\left(\bbox{\tilde{r}}-\bbox{\tilde{r}}^*(t)\right)^2-
    r_{TF}^2(t)\right)
\label{eq:Erg1}
\end{equation}
where the Thomas-Fermi radius $r_{TF}(t)$ satisfies
\begin{equation}
\ddot{r}_{TF}(t)-\bar{\omega}^2
 \left(\frac{r_{TF}^5(0)}{r_{TF}^4(t)}-\frac{1-p}{2}r_{TF}(t)\right)=0\,.
\label{eq:4.34}
\end{equation}
It is not influenced by the center of mass motion $\bbox{\tilde{r}}^*(t)$. The
phase of the state vector $\tilde{\psi}$ expressed in terms of $r_{TF}(t)$ is
\begin{eqnarray}
\tilde{\phi} &=&
\left(-\frac{\hbar k^2}{4M}+\frac{1+p}{2\hbar}\tilde{V}(0)-
 \frac{\omega}{2}\right)t\nonumber\\
&& - \frac{M}{\hbar}\int_0^td\tau\,
 \left(\frac{r_{TF}^3(0)}{r_{TF}^3(\tau)}-1\right)+
  \frac{M}{2\hbar}\left(\bbox{\tilde{r}}-\bbox{\tilde{r}}^*(t)\right)^2
   \frac{d}{dt}\ln\frac{r_{TF}(t)}{r_{TF}(0)}\nonumber\\
&& +\frac{M}{\hbar}\dot{\bbox{\tilde{r}}^*}(t)\cdot\bbox{\tilde{r}}\,
\label{eq:Erg2}
\end{eqnarray}
where energy conservation of the center of mass motion led to the cancellation
of some terms. Thus, for $p<1$ the condensate remains bound, even in the
presence of the electromagnetic outcoupling field. Its center of mass
oscillates with the frequency $\sqrt{\frac{1-p}{2}\bar{\omega}}$ around the
new equilibrium position at $\tilde{z}=-\frac{1+p}{1-p}\,
\frac{g}{\bar{\omega}^2}$. The Thomas-Fermi radius also oscillates, for small
amplitudes with frequency $\sqrt{\frac{5}{2}(1-p)}\bar{\omega}$ around its new
equilibrium value (see fig.\ref{6})
\begin{equation}
 r_{TF}=r_{TF}(0)\left(\frac{2}{1-p}\right)^{1/5}\,.
\label{eq:4.36}
\end{equation}

For $p>1$ the condensate becomes antitrapped by the outcoupling field and is
driven apart by the magnetic field of the trap. Its center of mass trajectory
can be inferred from (\ref{eq:4.31}) by analytic continuation to $p<1$ as
\begin{equation}
\bbox{r}^*(t)=\frac{p+1}{p-1}\,\frac{g\bbox{e}_z}{\bar{\omega}^2}
 \left(1-\cosh\left(\sqrt{\frac{p-1}{2}}\right)\bar{\omega}t\right)+
  \frac{\hbar\bbox{k}}{M}\,\frac{\sinh\left(\sqrt{\frac{p-1}{2}}\bar{\omega}t
   \right)}{\sqrt{\frac{p-1}{2}}\bar{\omega}}\,.
\label{eq:4:37}
\end{equation}
Its Thomas-Fermi radius (see fig.\ref{7}) grows with acceleration $\frac{p+1}{2}\bar{\omega}^2r_{TF}(0)$ for small times
$\bar{\omega}t\ll1$
\begin{equation}
r_{TF}(t)=r_{TF}(0)\left(1+\frac{p+1}{4}\bar{\omega}^2t^2\right)\,,
\label{eq:4.38}
\end{equation}
and exponentially at large times
\begin{equation}
r_{TF}(t)\sim\exp\left(\sqrt{\frac{p-1}{2}}\bar{\omega}t\right)
\label{eq:4.39}
\end{equation}
provided the condensate has not yet spread beyond the parabolic part of the
trapping potential.

\section{Output}\label{sec:5}

We assume the outcoupling field is turned off rapidly when, in a Rabi cycle
nearly all atoms are in the antitrapped state. To determine their macroscopic
wave function afterwards we have to solve their Gross-Pitaevskii equation
again.

\begin{figure}
\centerline{
\epsfig{figure=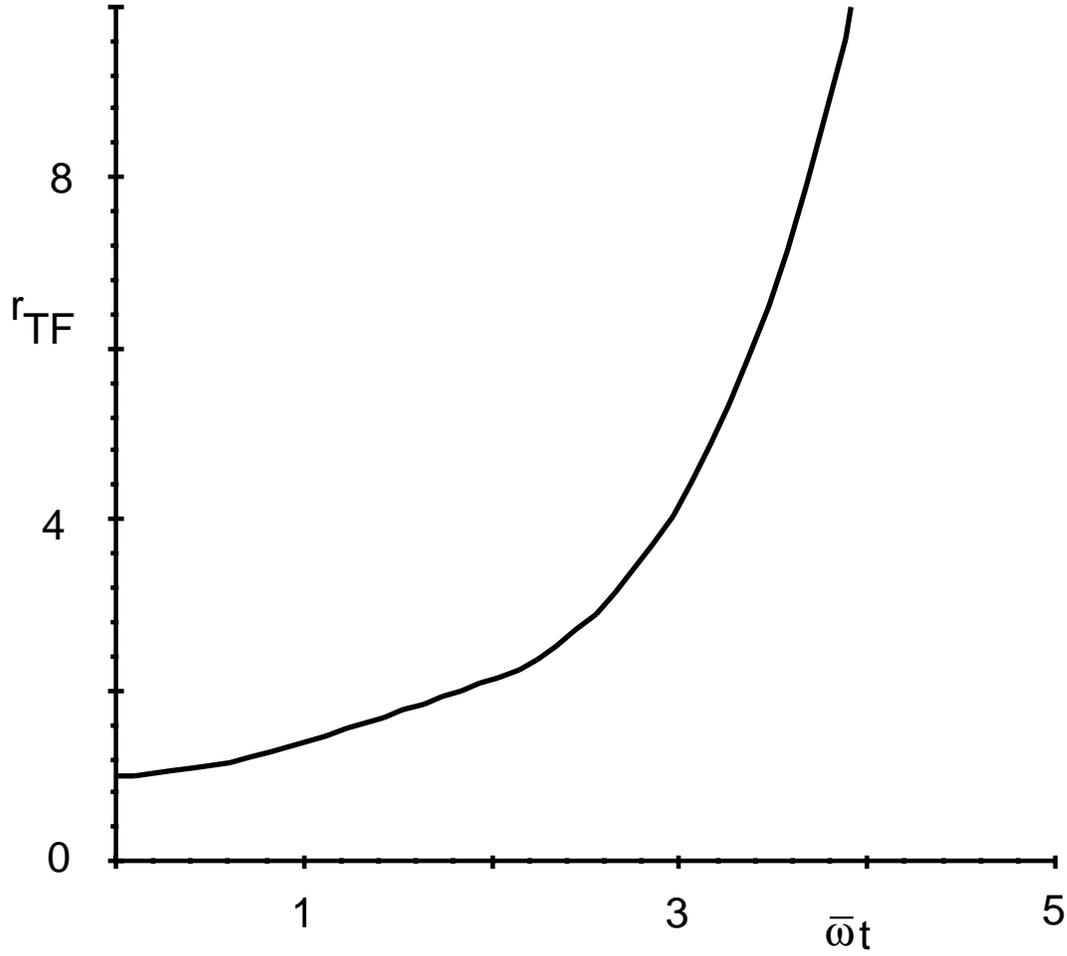}}
\caption{
The Thomas-Fermi radius $r_{FT}(t)$ in units of $r_{FT}(0)$ as a function of $\bar{\omega} t$
during $(\bar{\omega} t<2)$ and after $(\bar{\omega} t>2)$ Rabi oscillations
within a hyperfine manifold or between trapped and antitrapped states with
opposite potentials \label{8}
}
\end{figure}

{\bf 1. $F=1$ manifold}\\
      For outcoupling from the $F=1$ manifold
      \begin{equation}
      i\hbar\dot{\tilde{\psi}}_1=
      \left(-\frac{\hbar^2}{2M}\nabla^2-\frac{M}{2}\bar{\omega}^2\tilde{r}^2+
      2Mg\tilde{z}-\mu-2\tilde{V}(0)+2\hbar\omega+U_0|\tilde{\psi_1}|^2\right)
      \tilde{\psi}_1
      \label{eq:5.1}
      \end{equation}
      where atoms remaining in the $m=-1,0$ states have been neglected in the
      total density. This is justified if the outcoupling field is switched
      off at a time $T$ satisfying
      \begin{equation}
      \sqrt{2}\Omega T=(2n+1)\pi
      \label{eq:5.2}
      \end{equation}
      with integer $n$. To solve (\ref{eq:5.1}) we can use the same method as
      in section \ref{sec:4}. The center of mass $\bbox{\tilde{r}}^*(t)$ for
      $t>T$ satisfies
      \begin{equation}
      \ddot{\bbox{\tilde{r}}}^*=\bar{\omega}\bbox{\tilde{r}}^*-2g\bbox{e}_z\,.
      \label{eq:5.3}
      \end{equation}
      With the apropriate initial conditions at time $t=T$
      it is given by
      \begin{equation}
      \bbox{\tilde{r}}^*(t)=
      \frac{2g\bbox{e}_z}{\bar{\omega}^2}
      \left(1-\left(1+\frac{\bar{\omega}^2T^2}{4}\right)
      \cosh\bar{\omega}(t-T)-\frac{\bar{\omega}T}{2}
      \sinh\bar{\omega}(t-T)\right)\,,
      \label{eq:5.4}
      \end{equation}
      i.e. the combined action of gravity and the antitrapping potential
      pushes the center of mass of the outcoupled condensate vertically
      downwards. Meanwhile its Thomas-Fermi radius keeps expanding according
      to the equation of motion
      \begin{equation}
      \ddot{r}_{TF}(t)-\bar{\omega}^2
      \left(\frac{{r}_{TF}^5(0)}{{r}_{TF}^4(t)}+{r}_{TF}(t)\right)=0
      \label{eq:5.5}
      \end{equation}
\vspace{0.5cm}
      which corresponds to eq.~(\ref{eq:4.34}) with $p=3$. It has to be solved
      with initial conditions at time $t=T$, ${r}_{TF}={r}_{TF}(T)$ and
     \begin{equation}
     \dot{{r}}_{TF}(T)=\sqrt{\frac{2}{3}}\bar{\omega}{r}_{TF}(0)
     \left(1-\frac{{r}_{TF}^3(0)}{{r}_{TF}^3(T)}\right)^{1/2}
     \label{eq:5.5a}
     \end{equation}
     following from the solution for $t\le T$ of eq.~(\ref{eq:4.21a}). The solution for ${r}_{TF}(T)$ is   plotted
     in fig.\ref{8}. For times $\bar{\omega}t\gg1$ but still small enough so that
     the atomic cloud has not left the parabolic part of the trap, the cloud's
     center of mass coordinate and its radius both grow exponentially with the
     same rate $\bar{\omega}$. The density and phase of the outcoupled wave
     function are then given by eqs.~(\ref{eq:Erg1}), ~(\ref{eq:Erg2}), where
     eq.~(\ref{eq:Erg2}) has to be taken for $p=3$. Fig.\ref{8 Einschub1} gives
     a plot of the number density $|\psi|^2$ in the antitrapped stated.
     Plotted is $|\psi|^2$ in units of $15N/8\pi r_{TF}^3(0)$ as a function of
     the radial coordinate in the center of mass frame in units of $r_{TF}(0)$
     and $\Omega t$. In order to portray the influence of the Rabi-oscillations and the motion in the trap potential together the ratio $\Omega/\bar{\omega}$ in this and in the following two plots  has not been taken as large as would be desirable for the clear separation of time-scales we have assumed. In this plot $\bar{\omega}/\Omega=0.1$,
     and the duration of the radio-frequency pulse is $\bar{\omega}T=2$.

{\bf 2. $F=2$ manifold}\\
      In the limit $\Delta/\Omega\to0$ a complete transfer of
     the occupation probability from the trapped $m_F=2$ state to a single
     antitrapped state  (cf. fig.\ref{4}) occurs at the times
     $\Omega T=(2n+1)\pi$ when  all
     atoms are in the antitrapped $m_F=-2$ state. We can take over the
     preceding analysis with minor modifications.    The Gross-Pitaevskii     equation of the atoms in the $m_F=-2$ state
    \begin{equation}
    i\hbar\dot{\tilde{\psi}}_{-2}=
    \left(
    -\frac{\hbar^2}{2M}\nabla^2-\frac{M}{2}\bar{\omega}^2\tilde{r}^2+
     2Mg\tilde{z}-\mu-2\tilde{V}(0)+4\hbar\omega+U_0|\tilde{\psi}_{-2}|^2
    \right)\tilde{\psi}_{-2}
    \label{eq:5.7}
    \end{equation}
    takes the same form as (\ref{eq:5.1}) if there the replacement
    $\hbar\omega\to2\hbar\omega$ is made. The result for the density in the $m_F=-2$ state is plotted
    in fig.\ref{8 Einschub2}.

{\bf 3. Microwave and Raman outcoupling}\\
     In the limit $\Delta\ll\Omega$ all atoms in the condensate are in the
     antitrapped state $m=2$ (for $p>0$) at times $\Omega T=\frac{2n+1}{2}\pi$.
     The subsequent evolution of $\psi_2$ is governed by
     \begin{eqnarray}
     i\hbar\frac{\partial\tilde{\psi}_2}{\partial t}&=&
     \Bigg(-\frac{\hbar^2}{2M}\nabla^2-\frac{p}{2}M\bar{\omega}^2\tilde{r}^2+
     (1+p)Mg\tilde{z}+U_0|\psi_2|^2\nonumber\\
     &&\\
     \label{eq:5.8}
     &&\quad -\mu+\frac{\hbar^2 k^2}{2M}+\hbar\omega-(1+p)\tilde{V}(0)\Bigg)
     \tilde{\psi}_2\,.\nonumber
     \end{eqnarray}
     The center of mass satisfies the equation of motion
     \begin{equation}
     \ddot{\tilde{\bbox{r}}}^*=p\bar{\omega}^2\tilde
     {\bbox{r}}^*-(1+p)g\bbox{e}_z
      \label{eq:5.8a}
     \end{equation}
      with initial conditions following from eq.(\ref{eq:4.31}).  The
      Thomas-Fermi radius satisfies
     \begin{equation}
     \ddot{r}_{TF}(t)-\bar{\omega}^2
     \left(\frac{{r}_{TF}^5(0)}{{r}_{TF}^4(t)}+p{r}_{TF}(t)\right)=0
     \label{eq:5.9}
     \end{equation}
     with initial conditions ${r}_{TF}(T)$ and
     \begin{equation}
     \dot{r}_{TF}(T)=\bar{\omega}{r}_{TF}(0)
     \sqrt{\frac{2}{3}\left(1-\frac{{r}_{TF}^3(0)}{{r}_{TF}^3(T)}\right)+
     \frac{p-1}{2}\left(\frac{{r}_{TF}^2(T)}{{r}_{TF}^2(0)}-1\right)}
     \label{eq:5.10}
     \end{equation}
     following from eq.~(\ref{eq:4.34}). The solution for $p=1$ is shown in
     fig.\ref{8}. The density $|\psi_2|^2$ is
     then  given
     in terms of $\bbox{\tilde{r}}^*(t)$ and ${r}_{TF}(t)$ by
     eq.~(\ref{eq:Erg1}). For the purpose of generating an output pulse with
     a small Thomas-Fermi radius the case $p=0$ is of particular advantage.
     The number density in the excited state for this case is plotted in
     fig.\ref{8 Einschub3}. In this plot it is assumed that
     $\Omega/\bar{\omega}=5.91$ and $\Omega T=15\pi/2$. Finally, the phase is
     obtained as an expression  similar to eq.~(\ref{eq:Erg2}), namely
     \begin{eqnarray}
     \tilde{\phi}(\bbox{\tilde{r}},t) &=&
     \left(-\frac{\hbar^2 k^2}{2M}+(1+p)\tilde{V}(0)-\hbar\omega\right)
     \frac{t-T/2}{\hbar}\nonumber\\
     &&
     -\frac{\mu}{\hbar}\int_0^td\tau\
     \left(\frac{{r}_{TF}^3(0)}{{r}_{TF}^3(\tau)}-1\right)+
     \frac{M}{2\hbar}\left(\bbox{\tilde{r}}-\bbox{\tilde{r}}^*(t)\right)^2
     \frac{d}{dt}\ln\frac{{r}_{TF}(t)}{{r}_{TF}(0)}\\
     \label{eq:5.13}
     &&
     +\frac{M}{\hbar}
     \dot{\bbox{\tilde{r}}^*}(t)\cdot\bbox{\tilde{r}}\nonumber
     \end{eqnarray}
     where ${r}_{TF}^3(\tau)$ in the time integral has to be taken as solution
     of eq.~(\ref{eq:5.5}) for $0\le\tau\le T$ and as solution of
     eq.~(\ref{eq:5.9}) for $T\le\tau\le t$.

\section{Conclusion}

In this work we have studied outcoupling from magnetically trapped
Bose-Einstein condensates via electromagnetic fields resonantly coupling
either all the magnetic sublevels in the hyperfine manifold of the trapped
state or the trapped state and a state in a different hyperfine manifold.
Simplifying assumptions we made were (i) isotropic trap potentials, (ii)
equally spaced magnetic sublevels, i.e. neglect of quadratic Zeeman shifts,
(iii) interaction of the particles via their total density, (iv) the
validity of a dynamical version of the Thomas-Fermi approximation for the condensate, in which the kinetic energy per atom is small compared to the chemical potential and (for $\hbar=1$) the Rabi-frequency, and (v) the negligibility of thermal excitations. With
these assumptions we have given an analytical treatment of the problem of
strong outcoupling, where everywhere in the condensate the Rabi frequency
dominates over the space-dependent detuning, which is at most of the order of the
chemical potential. The equations of motion derived in that limit
(eqs.(\ref{eq:4.12}), (\ref{eq:4.13}), and (\ref{eq:4.15}) for rf-outcoupling,
eqs.(\ref{eq:4.12}), (\ref{eq:4.13}), (\ref{eq:4.28}), (\ref{eq:4.29}) for microwave and Raman oucoupling) could be solved exactly. For its quantitative validity our analytical treatment
depends crucially on the assumptions (ii), (iii), (iv), and (v), but we expect
that qualitatively our results should even be applicable if these
assumptions are not well satisfied. On the other hand the method we use can be
generalized to anisotropic harmonic potentials, so that assumption (i) could be relaxed,
but at the cost of more tedious analytical expressions.
Our method of solution expresses the densities and the phases of the
time-dependent macroscopic wavefunctions in the coupled states in terms of the
center of mass coordinate and the Thomas-Fermi radius of the condensate and theoutcoupled state and provides the ordinary second order differential equations
satisfied by these quantities.

The center of mass motion is in all cases very simple. It is, of course, independent of the interaction between the atoms and driven by the momentum transfer from the outcoupling field to the atoms
in the case of the Raman outcoupling, by gravity, and by the time-averaged trapping potential seen by the atoms. The latter vanishes for resonant rf-outcoupling within the hyperfine manifold of the condensate, i.e. the center of mass in this case is freely falling after the outcoupling field is switched on. If the outcoupling field is off-resonance, and in the cases of microwave or Raman outcoupling, the time-averaged trapping potential is, in general, non-zero (cf. eq.(\ref{eq:4.31}) for the center of mass motion in this case)
and leads either to a residual harmonic binding of the center of mass to the trap or to exponential repulsion from it.

\begin{figure}\centerline{
\epsfig{figure=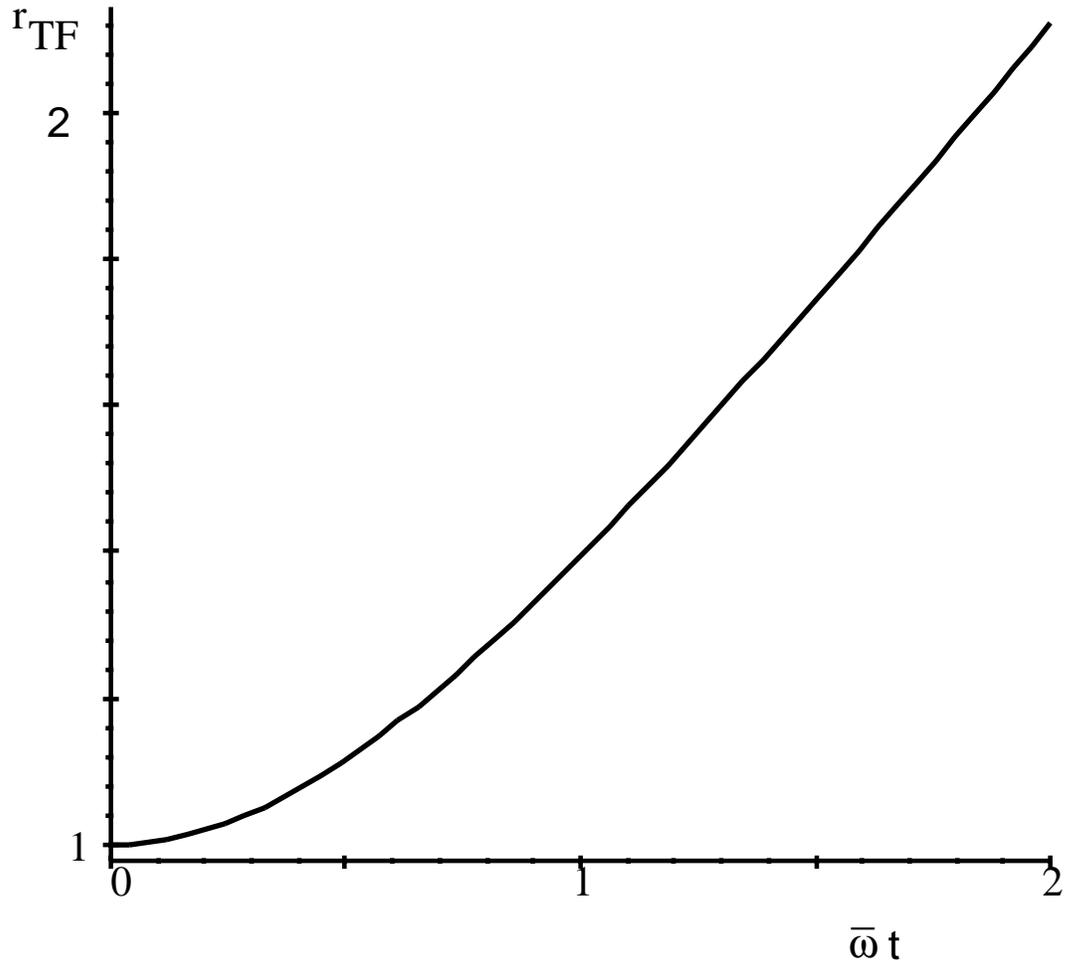}}
\caption{
The Thomas-Fermi radius  $r_{FT}(t)$ in units of $r_{FT}(0)$ as a function of $\bar{\omega} t$
during Rabi oscillations within a hyperfine-manifold \label{5}
}
\end{figure}

\begin{figure}
\centerline{
\epsfig{figure=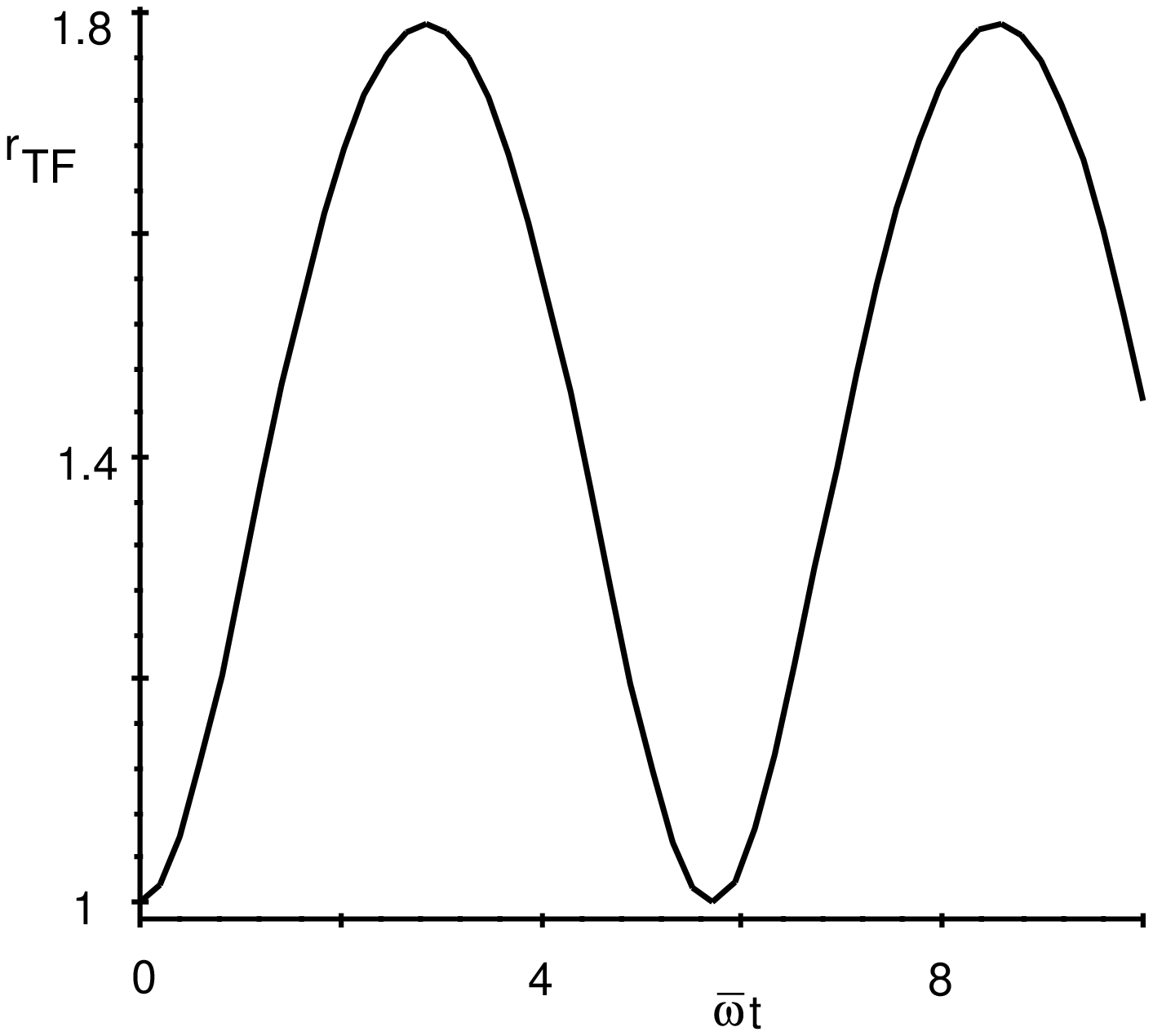}}
\caption{
The Thomas-Fermi radius $r_{FT}(t)$ in units of $r_{FT}(0)$ as a function of $\bar{\omega} t$
during Rabi oscillations between a trapped and a weakly anti-trapped state (p=1/2)
\label{6}
}
\end{figure}

\begin{figure}
\centerline{
\epsfig{figure=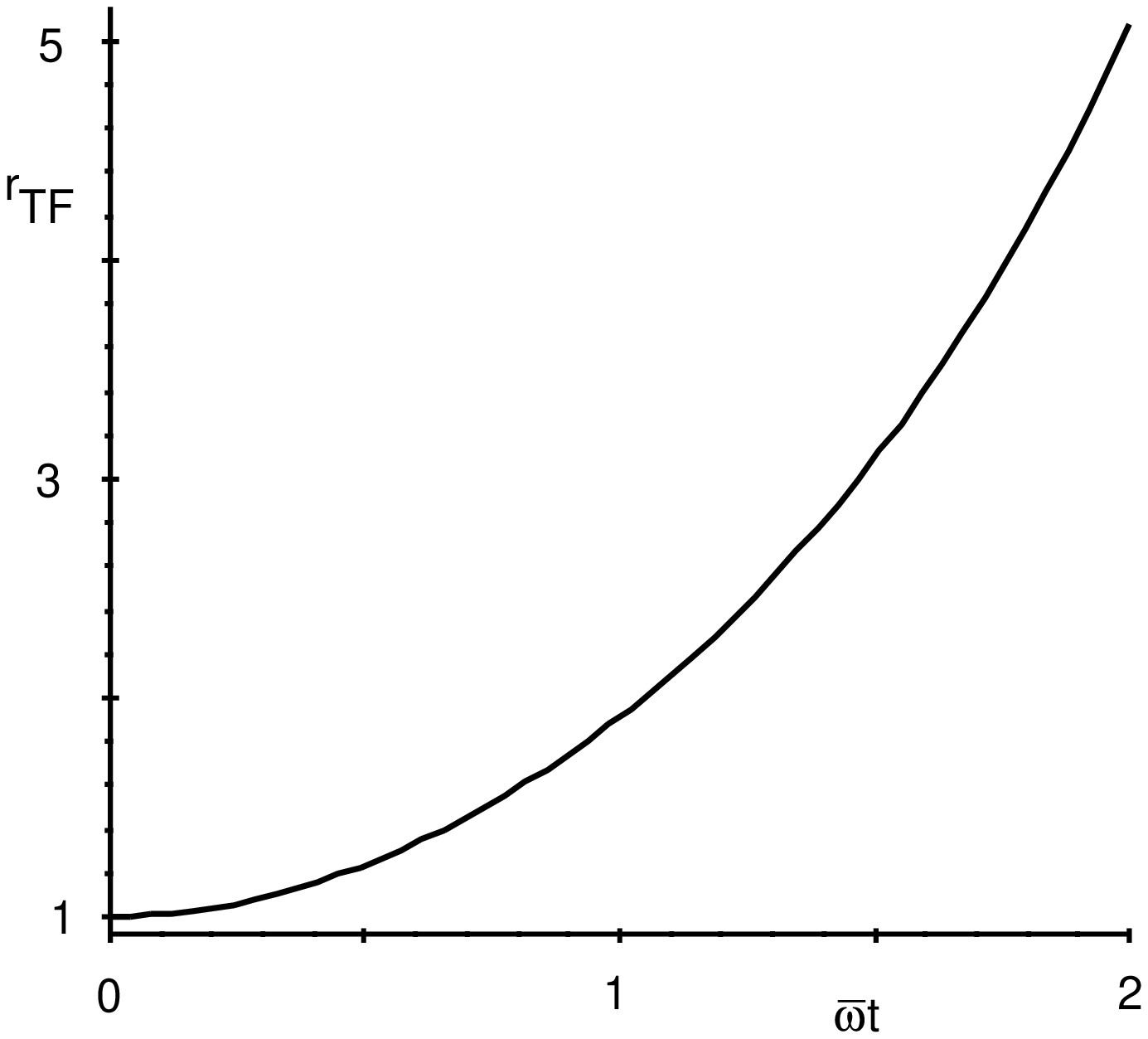}}
\caption{
The Thomas-Fermi radius  $r_{FT}(t)$ in units of $r_{FT}(0)$ as a function of $\bar{\omega} t$
during Rabi oscillations between a trapped and a strongly anti-trapped state (p=3)
\label{7}
}
\end{figure}

The second parameter characterizing the condensate in the outcoupling fields
is the time-dependent Thomas-Fermi radius. Its equation of motion (eq.(\ref{eq:4.21a}) while the outcoupling fields are on, eq.(\ref{eq:5.5}) or (\ref{eq:5.9}) after the outcoupling fields are switched off) are still very simple and can be solved analytically up to quadratures. In figs.\ref{5}-\ref{8} the three different types of solutions are portrayed.

Fig.\ref{5} shows the expansion of the Thomas-Fermi radius for the resonant rf-outcoupling within the hyperfine manifold of the condensate, which is the solution of eq.(\ref{eq:4.21a}). An initial phase of accelerated expansion due to the mutual repulsion of the atoms is followed by a purely ballistic expansion with constant velocity given by eq.(\ref{eq:4.23}). If the outcoupling field is switched off the density of atoms in the antitrapped states enter a phase of expontial expansion. The total evolution of $r_{TF}(t)$ for this case is shown in fig.\ref{8} while the complete evolution of the density in the antitrapped state is seen in fig.\ref{8 Einschub1} for the $F=1$
manifold and in fig.\ref{8 Einschub2} for the $F=2$ manifold. The only differencebetween
these two cases turns out to be the pattern of the Rabi-oscillations, seen directly in figs.\ref{3}, \ref{4}.
Fig.\ref{7} shows the expansion of the Thomas-Fermi radius for a case where the time-averaged potential seen by the atoms is repulsive. The expansion again starts out with constant acceleration and crosses over to an expontial
phase even while the outcoupling fields are on.
The third possibility for the evolution of $r_{TF}(t)$ is portrayed in fig.\ref{6}, which applies to the case where the time-averaged trapping potential seen by the atoms in the outcoupling field remains attractive.
The Thomas-Fermi radius oscillates in this case around its new enlarged equilibrium value either until  dissipation takes over (neglected throughout this work, which is justified if only a few oscillation cycles are considered)
or until the outcoupling fields are switched off. If this is done at a moment
where most atoms are in the outcoupled state, which may be untrapped or antitrapped, a ballistic or exponential
expansion follows the oscillations, respectively.
The complete evolution of the density of atoms outcoupled  to an untrapped state is shown in fig.\ref{8 Einschub3}.

It becomes clear from our treatment
that external electromagnetic fields inducing rapid Rabi oscillations are
excellent tools for suddenly changing or effectively turning off the trapping
potential, thereby exciting the center of mass and the dilation-compression
modes of the condensate. A further freedom in the possible changes of the
effective trapping potential is the introduction of a constant overall
detuning of the Rabi oscillations, if $\omega$ is chosen in such a way that nowhere
in the condensate the resonance condition is satisfied. Then only a small fraction of the atoms in the condensate is coupled out by each pulse of the external fields. The strong-coupling
limit in this case is reached if the effective Rabi frequency, including the detuning, is large compared to the variation of the detuning across the condensate. Our treatment, in
which resonance was assumed, may easily be extended to this case.

As long as the external electromagnetic
field is on, the condensate is not yet coupled out, unless the antitrapped
states to which it is coupled happens to be more strongly repelled by the
magnetic trap than the trapped states are attracted, which is the case $p>1$ in eq.(\ref{eq:3.11}). In all other cases, how many
atoms are finally coupled out depends crucially on the phase $\Omega T$ of
the Rabi cycle at the moment when the electromagnetic field is switched off. In
figs.\ref{8 Einschub1}-\ref{8 Einschub3} we have chosen the switch-off time $T$ in such
a way that the outcoupled state has its maximum occupation probability
precisely equal to or at least close to 1. A comparison of these figures, and of
the formulas to which they correspond, shows that outcoupling to an untrapped
state as in fig.\ref{8 Einschub3} rather than an antitrapped state as in
figs.\ref{8 Einschub1}, \ref{8 Einschub2} is highly preferrable to achieve a spatially confined,
narrow pulse, as the spatial width of the pulse in the antitrapped states is rapidly
broadened by the repelling trap potential. However, the coherence of the outcoupled
atom wave should permit a refocussing of the outcoupled pulse by
the techniques developed in atom optics.

We have followed the dynamics of the atoms only in the region of space where
they feel the parabolic part of the trap potential. In order to follow them
further as they leave the trap and finally become freely propagating atoms
it is necessary to make detailed assumptions about the trapping potential
far from the center of the trap. If this can be done for a given experiment
one may use semiclassical methods to solve the Schr\"odinger equation for the
atoms \cite{5}, eventually neglecting their interaction if the atomic beam hasbecome sufficiently dilute. This neglection is possible as soon as the $r_{TF}^{-4}$-term
in the equation of motion for $r_{TF}$ has become negligible. Our present
results for the region inside the trapping potential provides the initial or
boundary condition which is required for such a further analysis.

\noindent
{\Large\bf Acknowledgements}

We wish to thank Dr. Kalle-Antti Suominen for alerting us to an error in a preprint version of this work. This research was supported by the Marsden Fund of the Royal Society of New Zealand and the University of Auckland research committee. One of us (R.G) wishes to thank for the hospitality in the quantum optics group
at the University of Auckland and for the support by the Deutsche Forschungsgemeinschaft through the Sonderforschungsbereich 237 ``Unordnung
und gro{\ss}e Fluktuationen''.

\appendix
\section*{}
Here we record the needed analytic expressions for the Rabi oscillations in the various cases considered in this work.

{\bf F=1 manifold:}\\
The ansatz
\begin{equation}
   \tilde{\psi}=e^{-i\int_0^td\tau\ [\epsilon(\tilde{\bbox{r}},\tau )+\Delta(\tilde{\bbox{r}})]}|\psi(\tilde{\bbox{r}},t)|\varphi(\Omega t)\
    \label{eq:A1}
    \end{equation}
with $|\varphi |^2=1$ transforms eq.(\ref{eq:Os}) into
\begin{equation}
i\varphi'=
\left(\begin{array}{rcc}
-\delta & 1 & 0\\
1 & 0 & 1\\
0 & 1 & \delta\end{array}\right)\varphi
\label{eq:A2}
\end{equation}
with $\delta=\Delta/\Omega$, where $\varphi'$ is the derivative of $\varphi$
with respect to $\tau =\Omega t$. Its solution with initial condition
$\varphi_m=\delta_{m,1}$ is given by
\begin{equation}
\varphi = \frac{1}{2+\delta ^2}
\left(\begin{array}{c}
1+(1+\delta ^2)cos(\sqrt{2+\delta ^2}\tau)+i\delta\sqrt{2+\delta ^2}sin(\sqrt{2+\delta ^2}\tau)\\
\delta (1-cos(\sqrt{2+\delta ^2}\tau))-i\sqrt{2+\delta ^2}sin(\sqrt{2+\delta ^2}\tau)\\
-1+cos(\sqrt{2+\delta ^2}\tau)\end{array}\right).
\label{eq:A3}
\end{equation}
In the present work we need $\varphi (\Omega t)$ for $\delta =0$ and
\begin{equation}
(\bbox{\nabla}\varphi)_{\delta =0} = (\frac{\partial\varphi}{\partial\delta})_{\delta =0}
\bbox{\nabla}\delta\
\label{eq:A4}
\end{equation}
in order to evaluate the current velocity $(\hbar /M)Im[\varphi ^*\bbox{\nabla}\varphi ]_{\delta =0}$. Using (\ref{eq:A3}) we obtain
\begin{equation}
(\frac{\partial\varphi}{\partial\delta})_{\delta =0}=
\frac{1}{2}\left(\begin{array}{c}
i\sqrt{2}sin(\sqrt{2}\Omega t)\\
1-cos(\sqrt{2}\Omega t)\\
0\end{array}\right)
\label{eq:A5}
\end{equation}
and
\begin{equation}
Im[\varphi ^*\bbox{\nabla}\varphi ]_{\delta =0}=\frac{\sqrt{2}}{2\Omega}sin(\sqrt{2}\Omega t)\bbox{\nabla}\Delta\
\label{eq:A51}
\end{equation}

{\bf F=2 manifold:}\\
The ansatz
\begin{equation}
   \tilde{\psi}=e^{-i\int_0^td\tau\ [\epsilon(\tilde{\bbox{r}},\tau )+2\Delta(\tilde{\bbox{r}})]}|\psi(\tilde{\bbox{r}},t)|\varphi(\Omega t)\
    \label{eq:A6}
    \end{equation}
with $|\varphi |^2=1$ transforms eq.(\ref{eq:F21}) into
\begin{equation}
i\varphi'=
\left(\begin{array}{rrccc}
-2\delta & 1 & 0 & 0 & 0 \\
1 & -\delta & \sqrt{\frac{3}{2}} & 0 & 0 \\
0 & \sqrt{\frac{3}{2}} & 0 & \sqrt{\frac{3}{2}} & 0 \\
0 & 0 & \sqrt{\frac{3}{2}} & \delta & 1 \\
0 & 0 & 0 & 1 & 2\delta \end{array}\right)\varphi\
\label{eq:A7}
\end{equation}
with the same notation as in (\ref{eq:A2}). The solution with initial condition
$\varphi_m (0)=\delta_{m,2}$
is given by
\begin{eqnarray}
\varphi (\tau) = \varphi (0) &+& C_1[cos(\lambda_1\tau)+cos(\lambda_2\tau)-2]+C_2\frac{cos(\lambda_1\tau)-cos(\lambda_2\tau)}{\lambda_1^2-\lambda_2^2}\nonumber \\
\label{eq:A8}\\
&+& iC_3[\lambda_1sin(\lambda_1\tau)+\lambda_2sin(\lambda_2\tau)]+iC_4\frac{\lambda_1sin(\lambda_1\tau)-\lambda_2sin(\lambda_2\tau)}{\lambda_1^2-\lambda_2^2}\nonumber\
\end{eqnarray}
with the characteristic dimensionless frequencies
\begin{equation}
\lambda_{1}=2\sqrt{\delta^2+1}, \quad \lambda_{2}=\sqrt{\delta^2+1}\label{eq:A9}
\end{equation}
and the coefficient-vectors
\begin{eqnarray}
C_1 =
\frac{1}{16(\delta^2+1)^2}
\left(\begin{array}{c}
8\delta^4+16\delta^2+5\\
-6\delta\\-\sqrt{6}(2\delta^2-1)\\
6\delta\\
-3
\end{array}\right);
C_2 &=&
\frac{1}{16(\delta^2+1)}
\left(\begin{array}{c}
24\delta^4-9\\
-6\delta (8\delta^2+3)\\
3\sqrt{6}(6\delta^2+1)\\
-30\delta\\
15
\end{array}\right)\nonumber\\
\label{eq:A10}\\
C_3 =
\frac{1}{16(\delta^2+1)^2}
\left(\begin{array}{c}
2\delta(2\delta^2+5)\\
4\delta^2-5\\-3\sqrt{6}\delta\\
3\\
0\end{array}\right);
C_4 &=&
\frac{1}{16(\delta^2+1)}
\left(\begin{array}{c}
6\delta (2\delta^2-3)\\
-9(4\delta^2-1)\\
15\sqrt{6}\delta\\
-15\\
0\end{array}\right)\nonumber\
\end{eqnarray}
The special case $\delta =0$ gives $\lambda_1=2$, $\lambda_2=1$ and leads to eq.(\ref{eq:3.10}). In the present work we also need $(\bbox{\nabla}\varphi)_{\delta =0}$ and hence $(\frac{\partial\varphi}{\partial\delta})_{\delta =0}$. Using (\ref{eq:A8})-(\ref{eq:A10}) we obtain
\begin{equation}
(\frac{\partial\varphi}{\partial\delta})_{\delta =0}=
\left(\begin{array}{c}
\frac{i}{2}(sin(2\Omega t)+sin(\Omega t))\\
-\frac{3}{4}(cos(2\Omega t)-1)\\
\frac{i\sqrt{6}}{4}(sin(2\Omega t)-2sin(\Omega t))\\
-\frac{1}{4}cos(2\Omega t)+cos(\Omega t)-\frac{3}{4}\\
0
\end{array}\right)
\label{eq:A11}
\end{equation}
and
\begin{eqnarray}
Im[\varphi ^*\bbox{\nabla}\varphi ]_{\delta =0} &=& \frac{2}{\Omega}sin(\Omega t)\bbox{\nabla}\Delta\
\label{eq:A12}
\end{eqnarray}

{\bf Two-level case:}\\
The ansatz
\begin{equation}
   \tilde{\psi}=e^{-i\int_0^td\tau\ [\epsilon(\tilde{\bbox{r}},\tau )+\Delta(\tilde{\bbox{r}})/2]}|\psi(\tilde{\bbox{r}},t)|\varphi(\Omega t)\
    \label{eq:A13}
    \end{equation}
leads to
\begin{equation}
\varphi (\tau) = \frac{1}{\sqrt{\delta ^2+4}}
\left(\begin{array}{c}
\sqrt{\delta ^2+4}cos(\sqrt{1+\delta ^2/4}\tau)+i\delta sin(\sqrt{1+\delta ^2/4}\tau)\\
-2i sin(\sqrt{1+\delta ^2/4}\tau)\end{array}\right)
\label{eq:A14}
\end{equation}
from which we can extract the result for
\begin{equation}
Im[\varphi ^*\bbox{\nabla}\varphi ]_{\delta =0} = \frac{1}{4\Omega}sin(2\Omega t) \bbox{\nabla}\Delta\
\label{eq:A15}
\end{equation}

\begin{figure}
\caption{
\protect$|\psi_1|^2$ for outcoupling from the $F=1$ manifold in units of
\protect$15N/8\pi r^3_{TF}(0)$ as function of the radial coordinate in the
center of
mass frame in units of \protect$r_{TF}(0)$ and time in units
\protect$\Omega^{-1}$ for the
choice of parameters  \protect$\bar{\omega}/\Omega=0.1$ and
\protect$\Omega T=20$ \label{8 Einschub1}
}
\end{figure}

\begin{figure}
\caption{
$|\psi_{-2}|^2$ for outcoupling from the $F=2$ manifold plotted as in fig.\ref{8 Einschub1} for
$\bar{\omega}/\Omega=0.1$ and $\Omega T=22$ \label{8 Einschub2}
}
\end{figure}

\begin{figure}
\caption{
$|\psi_2|^2$ for outcoupling from a trapped state to an untrapped state
$(p=0)$ in a different hyperfine manifold plotted as in fig.\ref{8 Einschub1} for
$\bar{\omega}/\Omega=0.169$ and $\Omega T=15\pi/2$ \label{8 Einschub3}
}
\end{figure}


\begin{thebibliography}{99}
\bibitem{1} M.-O.~Mewes, M.~R.~Andrews, D.~M.~Kurn, D.~S.~Durfee,
            C.~G.~Townsend, and W.~Ketterle,
            {\it Phys.~Rev.~Lett. \bf 78}, 585 (1997)
\bibitem{2} L.~Deng, E.~Hagley, M.~Kozuma, R.~Lutwak, Yu.~Ovchinnikov,
            J.~Wen,   K.~Helmerson, W.~O.~Phillips, S.~L.~Rolston,
            paper  presented at IQEC 1998
\bibitem{3} R.~J.~Ballagh, K.~Burnett, and T.~F.~Scott,
            {\it Phys.~Rev.~Lett. \bf 78}, 1607 (1997)
\bibitem{3a} W.~Zhang and D.~F.~Walls,
            {\it Phys.~Rev. \bf A57}, 1248 (1998);
            {\it Phys.~Rev. \bf A58}, 4248 (1998) (Erratum)
\bibitem{4} H.~Steck, Thesis, Ludwig-Maximilian-Universit\"at M\"unchen 1997;
            H.~Steck, M.~Naraschewski and H.~Wallis,
            {\it Phys.~Rev.~Lett. \bf 80 },1 (1998)
\bibitem{HMS} G.~M.~Moy, J.~J.~Hope and C.~M.~Savage,
            {\it Phys.~Rev. \bf A }, ()
\bibitem{5} M.~W.~Jack, M.~Naraschewski, M.~J.~Collet, and D.~F.~Walls,
            {\it Phys.~Rev. \bf A }, ()
\bibitem{5a} D.~A.~W.~Hutchinson,
            cond-mat/9811129
\bibitem{6} I.~Bloch, Th.~W.~H\"ansch and T.~Esslinger, cond-mat/9812258
\bibitem{7} Y.~Kagan, E.~L.~Surkov, and G.~Shlyapnikov,
            {\it Phys.~Rev. \bf A54}, R1753 (1996);
            {\it Phys.~Rev. \bf A55}, R18 (1997)
\bibitem{8} Y.~Castin and R.~Dum,
            {\it Phys.~Rev.~Lett. \bf 77}, 5115 (1996)
\bibitem{F1} N.~V.~Vitanov and K.~-A.~Suominen, {\it Phys.~Rev. \bf A56}, R4377
(1997)
\bibitem{F2} K.~-A.~Suominen, E.~Tiesinga and P.~S.
~Julienne, {\it Phys.~Rev. \bf A58}, 3983 (1998)
\bibitem{TF1} V.~V.~Goldman, I.~F.~Silvera, and A.~J.~Leggett, {\it Phys.~Rev.   \bf B24}, 2870 (1981)
\bibitem{TF2} D.~A.~Huse and E.~D.~Siggia, {\it J.~Low Temp.~Phys.\bf 46}, 137 (1982)
\end{thebibliography}
\end{document}